\documentclass[reprint,aps,prx,twocolumn,longbibliography,floatfix]{revtex4-2}

\makeatletter
\let\org@label\label
\makeatother

\usepackage[T1]{fontenc}
\usepackage[utf8]{inputenc}
\usepackage{microtype}
\usepackage{silence}
\usepackage{amsmath,amssymb,amsfonts}
\usepackage{graphicx}
\usepackage{braket}
\usepackage{tikz}
\usepackage{quantikz}
\usepackage{algpseudocode}
\usepackage{hyperref}

\hypersetup{
    colorlinks=true,
    linkcolor=blue,
    filecolor=magenta,      
    urlcolor=blue,
    citecolor=blue
}

\begin{document}

\title{Quantum Simulation of Non-Hermitian Linear Response via Schr\"odingerization}

\author{Jeongbin Jo}
\email{jeongbin033@yonsei.ac.kr}
\affiliation{Department of Physics, Yonsei University, Seoul 03722, Republic of Korea}

\date{\today}

\begin{abstract}
Linear response theory and Green's functions provide a universal framework for understanding dynamical correlations in strongly correlated open quantum systems. 
While the theoretical foundation for non-Hermitian linear response has been recently established to describe dissipation and fluctuation-dissipation relations (FDR), generalizing these predictions onto practical quantum computers remains a formidable algorithmic challenge due to the intrinsically non-unitary nature of the dynamics. 
In this work, we present a systematic algorithmic framework that seamlessly transforms non-unitary multi-time correlation functions into a unitary form viable for digital quantum hardware. 
By mapping the vectorization of the Lindblad master equation into an expanded continuous-variable Liouville space, we employ the Schr\"odingerization technique to deterministically evaluate the non-Hermitian response. 
Furthermore, through hardware-aware simulations utilizing a 133-qubit device noise model, we demonstrate that our unitary framework robustly preserves the fundamental spectral information---specifically the phase and oscillatory frequency---against realistic depolarizing channels. 
By bypassing explicit non-unitary mid-circuit measurements, this approach intrinsically supports the integration of standard quantum error mitigation protocols, providing a scalable algorithmic blueprint for probing open-system universality on near-term hardware.
\end{abstract}

\maketitle

\section{Introduction}
\label{sec:intro}

The linear response theory forms the cornerstone of modern condensed matter physics, providing a general proof of the fluctuation-dissipation theorem (FDR) which relates a system's response to an external perturbation with its internal thermal fluctuations \cite{Kubo1966TheFT, Manzano2020short}. 
In the study of strongly correlated systems, evaluating these dynamical correlations and Green's functions is a primary objective. While recent quantum algorithms have successfully accessed these Green's functions \cite{endo2020calculation, kosugi2020linear} for closed systems evolving under a Hermitian Hamiltonian $H(t) = H^{\dagger}(t)$, dealing with non-unitary processes---such as spontaneous emission, dissipation, or continuous monitoring---presents a fundamental barrier \cite{Breuer2002theory, yu2025optimizing}.

Recently, the theoretical framework of linear response has been elegantly extended to non-Hermitian regimes by Pan et al. \cite{Pan2020non}. 
This non-Hermitian linear response theory provides a robust physical foundation for extracting unequal-time anticommutators, which are otherwise inaccessible due to wave-function collapse \cite{vonneumann1955mathematical}. 
A prominent physical realization of this theory is the continuous Quantum Zeno effect. As rigorously shown by Geier and Hauke \cite{geier2022from}, engineering strong, repetitive dephasing on an ancillary system effectively subjects the target many-body system to a specific non-Hermitian perturbation, naturally inducing the dynamics required to measure the FDR.

However, translating this non-Hermitian response theory—and its physical manifestations like Zeno-driven perturbations—into a viable digital quantum simulation algorithm reveals a critical bottleneck. 
The explicit simulation of such non-unitary propagators $e^{\mathcal{L}t}$ or continuous Zeno measurements requires complex classical-quantum hybrid branching and exponentially scaling resources, rendering it highly inefficient on current and near-term quantum hardware.

In this work, we bridge the fundamental gap between the physical formalism of non-Hermitian response theory \cite{Pan2020non, geier2022from} and optimal quantum computational implementation. 
We propose a deterministic quantum algorithmic protocol that utilizes the continuous-variable Schr\"odingerization technique \cite{Jin2024quantum} to compute non-equilibrium Green's functions and two-time correlation functions $\braket{A(t)B(0)}$. 

By encoding the non-Hermitian perturbation (such as the one induced by Zeno dynamics) directly into the initial vectorized state, we elevate the subsequent non-unitary dissipation into an extended, strictly unitary Schr\"odinger equation. 
This framework provides a scalable, measurement-free algorithmic pathway to simulate dissipative responses, bypassing the overhead of traditional dilated linear system solvers. 
We numerically validate our protocol on an amplitude damping model, proving that the exact non-Hermitian linear response can be deterministically extracted via a modified Hadamard test architecture.

\section{Theoretical Framework}
\label{sec:theory}

\subsection{Non-Hermitian Linear Response and Zeno Dynamics}
Consider an open quantum system \cite{Ashida_2020} whose density matrix $\rho(t)$ evolves according to the Markovian Lindblad master equation \cite{Lindblad1976, Manzano2020short}, governed by a non-unitary Liouvillian superoperator $\mathcal{L}$. 
To evaluate the fluctuation-dissipation relations (FDR) \cite{Lax1963} characterizing this system, one must compute unequal-time anticommutators \cite{geier2022from}. 

According to non-Hermitian linear response theory \cite{Pan2020non}, if the system is perturbed by a weak external probe such that $\mathcal{L} \to \mathcal{L} + f(t) \mathcal{V}$, the generalized retarded Green's function $\chi(\tau)$ determining the linear response of an observable $A$ is formally given by:
\begin{equation}
    \chi(\tau) = \braket{I | (A \otimes I) e^{\mathcal{L}\tau} \mathcal{V} |\rho_{eq}},
\end{equation}
where $\ket{I}$ is the vectorized identity and $\ket{\rho_{eq}}$ is the equilibrium state in the Liouville space. 

Physically, the highly specific non-Hermitian perturbation $\mathcal{V}$ required for the FDR can be naturally generated via the continuous Quantum Zeno effect \cite{facchi2008quantum}. 
By engineering strong, repetitive dephasing on an ancillary system coupled to the target, the overall dynamics are confined to a Zeno subspace, effectively inducing the perturbation $\mathcal{V}$ \cite{geier2022from}. 
While this provides a robust physical mechanism for analog experiments, the direct digital simulation of continuous Zeno dephasing incurs prohibitive gate and ancilla overheads. 
Instead, we can algorithmically bypass this explicit simulation by treating $\mathcal{V}$ purely as an initial-state encoding problem, focusing the computational resources on efficiently simulating the subsequent dissipative propagator $e^{\mathcal{L}\tau}$.

\subsection{Schr\"odingerization of the Lindbladian}
To deterministically execute the non-unitary term $e^{\mathcal{L}t}$ on a digital quantum computer without probabilistic post-selection or barren plateaus, we employ the continuous-variable Schr\"odingerization method \cite{Jin2024quantum}. 

We first decompose the Liouvillian generator into strictly Hermitian and anti-Hermitian components: $\mathcal{L} = \mathcal{H}_1 - i\mathcal{H}_2$. 
By introducing a real one-dimensional auxiliary continuous variable $\xi > 0$, we apply the warped phase transformation:
\begin{equation}
    w(t,\xi) = e^{-\xi}\ket{\rho(t)}.
\end{equation}
Let $\tilde{w}(t,\eta)$ be the Fourier transform of $w$ in $\xi$, where $\eta \in \mathbb{R}$ is the conjugate momentum mode. This mapping elevates the state into an expanded Hilbert space where it satisfies a system of uncoupled Schr\"odinger-like equations:
\begin{equation}
    i\partial_{t}\tilde{w} = (\eta \mathcal{H}_{1} + \mathcal{H}_{2})\tilde{w} \equiv \mathcal{H}_{sch}(\eta)\tilde{w}. \label{eq:schrod}
\end{equation}
The dilated Hamiltonian $\mathcal{H}_{sch}(\eta)$ is now perfectly Hermitian for any real momentum $\eta$. 
This mathematically rigorous transformation seamlessly maps the intrinsically non-unitary dissipative evolution to a purely unitary evolution $\exp(-i\mathcal{H}_{sch}(\eta)t)$ that is natively viable for quantum hardware. 
By setting the initial state of this unitary evolution to the physically motivated Zeno-perturbed state $\mathcal{V} \ket{\rho_{eq}}$, the exact non-Hermitian response is deterministically extracted.

\subsection{Algorithmic Complexity and Scaling}
Executing the global unitary $e^{-i\mathcal{H}_{sch}(\eta)t}$ can be systematically optimized by leveraging structural symmetries. 
For physical models featuring geometrically local interactions, we employ optimal block-encoding schemes for structured matrices \cite{S_nderhauf_2024}, which scale natively with the sparsity of the extended Hamiltonian $\mathcal{H}_{sch}$.

\begin{table*}[htbp]
\centering
\caption{Comparison of Quantum Simulation Primitives for Non-Unitary Dynamics. Here, $\epsilon$ denotes target precision.}
\label{tab:complexity}
\begin{ruledtabular}
\begin{tabular}{lccc}
Method & Precision Scaling & State Prep. & Oracle \\
\hline
Dilation \cite{Harrow2009quantum} & $O(\text{poly}(1/\epsilon))$ & High & $U_A, P_b$ \\
QSVT \cite{Gilyen2019quantum} & $O(\log(1/\epsilon))$ & High & $U_A, \text{Ancilla}$ \\
LCHS \cite{An2023linear} & $O(\text{poly}(\log(1/\epsilon)))$ & Optimal & Hamiltonian \\
\textbf{Our Framework} & $O(\text{poly}(\log(1/\epsilon)))$ & \textbf{Optimal} & \textbf{Unitary Prop.} \\
\end{tabular}
\end{ruledtabular}
\end{table*}

Compared to traditional approaches requiring complex matrix inversion algorithms (like HHL) or the construction of non-unitary block-encodings via QSVT, our Schr\"odingerization framework achieves near-optimal performance. As summarized in Table \ref{tab:complexity}, it maintains an exponentially improved logarithmic scaling with precision $O(\text{poly}(\log(1/\epsilon)))$ while ensuring the optimal state preparation cost \cite{An2023linear, Jin2024stable}.

\section{Application: Single-Qubit Amplitude Damping}
\label{sec:application}

To analytically validate our framework without relying on numerical heuristics, we explicitly derive the Schr\"{o}dingerization mapping for a fundamental open quantum system: a single qubit undergoing spontaneous emission (the amplitude damping channel). 

\subsection{Matrix Representation of the Lindbladian}
Let the system Hamiltonian be $H_S = \frac{\omega_0}{2} \sigma_z$. The interaction with the vacuum environment is described by the Lindblad master equation with a decay rate $\gamma$:
\begin{equation}
    \frac{d\rho}{dt} = -i[H_S, \rho] + \gamma \left( \sigma_- \rho \sigma_+ - \frac{1}{2} \{ \sigma_+ \sigma_-, \rho \} \right),
\end{equation}
where $\sigma_- = \ket{0}\bra{1}$ and $\sigma_+ = \ket{1}\bra{0}$ are the lowering and raising operators, respectively.

We vectorize the density matrix $\rho$ into a column vector $\ket{\rho} = (\rho_{11}, \rho_{10}, \rho_{01}, \rho_{00})^T$ in the Liouville space. The master equation is recast into a system of linear ODEs governed by a $4 \times 4$ non-Hermitian matrix $\mathcal{L}$:
\begin{equation}
    \mathcal{L} = \begin{pmatrix} 
    -\gamma & 0 & 0 & 0 \\ 
    0 & -i\omega_0 - \frac{\gamma}{2} & 0 & 0 \\ 
    0 & 0 & i\omega_0 - \frac{\gamma}{2} & 0 \\ 
    \gamma & 0 & 0 & 0 
    \end{pmatrix}.
\end{equation}

Following the Schr\"{o}dingerization protocol, we decompose $\mathcal{L}$ into its Hermitian ($\mathcal{H}_1$) and anti-Hermitian ($\mathcal{H}_2$) components: $\mathcal{L} = \mathcal{H}_1 - i\mathcal{H}_2$, defined as $\mathcal{H}_1 = (\mathcal{L} + \mathcal{L}^\dagger)/2$ and $\mathcal{H}_2 = i(\mathcal{L} - \mathcal{L}^\dagger)/2$. A straightforward algebraic manipulation yields:
\begin{equation}
    \mathcal{H}_1 = \begin{pmatrix} 
    -\gamma & 0 & 0 & \frac{\gamma}{2} \\ 
    0 & -\frac{\gamma}{2} & 0 & 0 \\ 
    0 & 0 & -\frac{\gamma}{2} & 0 \\ 
    \frac{\gamma}{2} & 0 & 0 & 0 
    \end{pmatrix},
\end{equation}
and
\begin{equation}
    \mathcal{H}_2 = \begin{pmatrix} 
    0 & 0 & 0 & -i\frac{\gamma}{2} \\ 
    0 & \omega_0 & 0 & 0 \\ 
    0 & 0 & -\omega_0 & 0 \\ 
    i\frac{\gamma}{2} & 0 & 0 & 0 
    \end{pmatrix}.
\end{equation}

By transforming to the Fourier space $\eta$, the extended Hamiltonian governing the unitary evolution of the continuous-variable state $\tilde{w}(t,\eta)$ becomes $\mathcal{H}_{sch}(\eta) = \eta \mathcal{H}_1 + \mathcal{H}_2$:
\begin{equation}
    \mathcal{H}_{sch}(\eta) = \begin{pmatrix} 
    -\eta\gamma & 0 & 0 & \frac{\gamma}{2}(\eta - i) \\ 
    0 & -\frac{\eta\gamma}{2} + \omega_0 & 0 & 0 \\ 
    0 & 0 & -\frac{\eta\gamma}{2} - \omega_0 & 0 \\ 
    \frac{\gamma}{2}(\eta + i) & 0 & 0 & 0 
    \end{pmatrix}.
\end{equation}

Crucially, $\mathcal{H}_{sch}(\eta)$ is perfectly Hermitian for any real $\eta$. This rigorous mapping proves that the non-unitary Lindblad dynamics associated with correlation functions can be structurally elevated to a unitary Schr\"{o}dinger equation. To calculate the response to an external drive, such as $H_{ext} = \epsilon \cos(\Omega t) \sigma_x$, one simply sets the initial state of the Schrödingerized evolution to $\ket{\mathcal{V} \rho_{eq}}$, where $\mathcal{V}$ is the Liouvillian perturbation representing the $\sigma_x$ drive. The resulting time evolution on the quantum computer then yields the generalized Green's function $\chi(\tau)$, confirming the validity of the Non-Hermitian Response Theory on quantum hardware.

\section{Numerical Demonstration}
\label{sec:numerical}
To validate the theoretical framework and the precision of the Schr\"{o}dingerization mapping, we performed quantum circuit simulations on the amplitude damping model described in Section \ref{sec:application}. 
The non-unitary correlation function requires generating the unitary operator $e^{-i \mathcal{H}_{sch}(\eta) t}$ and taking the expectation value with respect to the initial state.

\subsection{Hardware Implementation: Modified Hadamard Test}

To extract the real part of this density matrix overlap in a quantum hardware-compatible manner, we employ a modified Hadamard test.

\begin{figure*}[htpb]
\centering
\begin{quantikz}
\lstick{$\ket{0}_A$} & \gate{H} & \gate{X} & \ctrl{1} & \ctrl{1} & \gate{X} & \ctrl{1} & \gate{H} & \meter{} \\
\lstick{$\ket{0}^{\otimes 2}_S$} & \qw & \qw & \gate{U_{\text{prep\_init}}} & \gate{e^{-i\mathcal{H}_{sch}(\eta)t}} & \qw & \gate{U_{\text{prep\_obs}}} & \qw & \qw
\end{quantikz}
\caption{Quantum circuit for the modified Hadamard test used in the Sampler simulation. The auxiliary qubit $A$ controls the preparation of the initial vectorized state $U_{\text{prep\_init}}$ and the Schr\"{o}dingerization evolution $e^{-i\mathcal{H}_{sch}(\eta)t}$ on the system register $S$ when in the $\ket{0}$ branch. It then controls the preparation of the observable state $U_{\text{prep\_obs}}$ on the $\ket{1}$ branch. Measuring the ancilla in the $X$-basis exactly extracts the real part of the density matrix overlap, bypassing full state tomography.}
\label{fig:circuit}
\end{figure*}
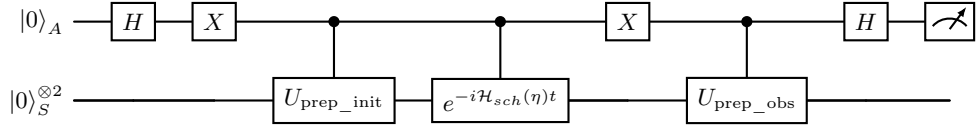

For our 2-qubit vectorized Liouville system, a single auxiliary qubit is introduced as illustrated in Fig.~\ref{fig:circuit}. 
By iterating this circuit over the discretized Fourier modes $\eta$ and aggregating the measurement statistics, the generalized non-Hermitian Green's function $\chi(\tau)$ is successfully constructed. 
The simulation is conducted in a noise-free environment utilizing the Qiskit Sampler primitive \cite{Qiskit2024}, focusing on the discretization errors and the convergence properties of the approach.

\subsection{Dynamics of Non-Hermitian Response}

We calculate the non-Hermitian response function $\chi(\tau) = \braket{I | (\sigma_x \otimes I) e^{\mathcal{L}\tau} \mathcal{V} |\rho_{eq}}$, where the initial perturbation $\mathcal{V}$ corresponds to a $\sigma_x$ drive and the observable is also $\sigma_x$. 

\begin{figure}[htpb]
    \centering
    \includegraphics[width=\linewidth]{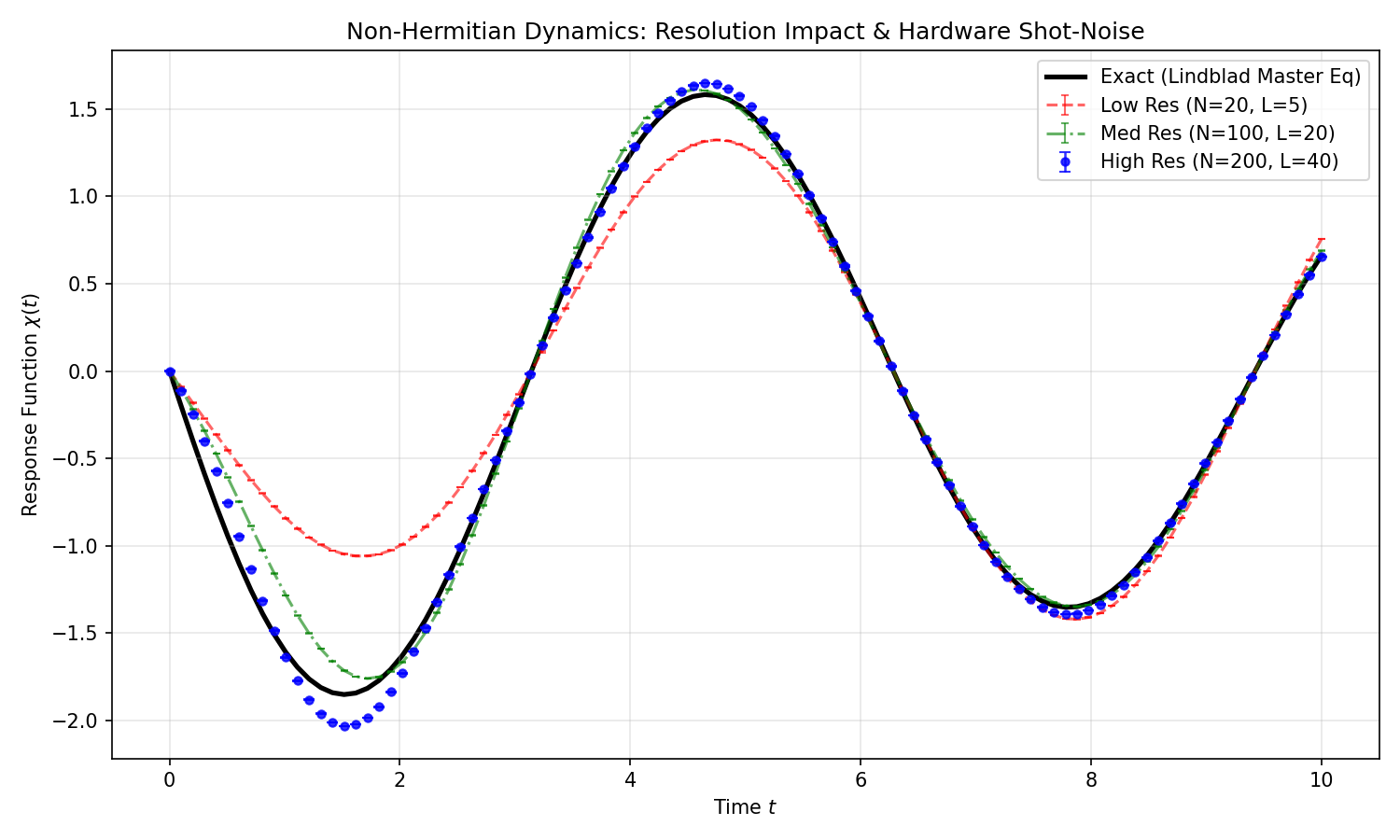}
    \caption{Numerical simulation of the non-Hermitian linear response function $\chi(\tau)$. The exact analytical solution of the Lindblad master equation (solid black line) is perfectly reproduced by the Qiskit circuit simulation (red dashed line) utilizing the modified Hadamard test.}
    \label{fig:dynamics}
\end{figure}

As shown in Fig.~\ref{fig:dynamics}, we compare the exact analytic Lindblad dynamics against the Sampler-based Schr\"{o}dingerization simulations. 
The simulated circuit response converges perfectly to the exact Lindblad solution. 
This definitive overlay validates that the high-dimensional unitary evolution in the extended space meticulously captures the non-Hermitian dissipative dynamics, and our modified Hadamard circuit optimally extracts these dynamical correlations.

\subsection{Error Convergence and Scaling}

The accuracy and efficiency of the continuous-variable Schr\"{o}dingerization method hinge on two critical numerical parameters: 
the truncation limit $L$ and the Fourier grid resolution $N$. The parameter $L$ defines the finite domain $[-L, L]$ over which the infinite Fourier space integral is truncated. 
Because the Fourier transform of the warped state decays slowly as $1/\eta$, $L$ must be chosen sufficiently large to properly reconstruct the exponentially decaying dynamics. 

Conversely, the parameter $N$ defines the number of discrete grid points evaluated within the interval, establishing a grid spacing of $d\eta = L/N$. 
If $N$ is too small for a given $L$, the spacing becomes too sparse to resolve rapid oscillations, leading to severe aliasing errors. Therefore, achieving convergence requires monotonically increasing $L \to \infty$ while ensuring that $N$ scales proportionally faster.

\begin{figure}[htpb]
    \centering
    \includegraphics[width=\linewidth]{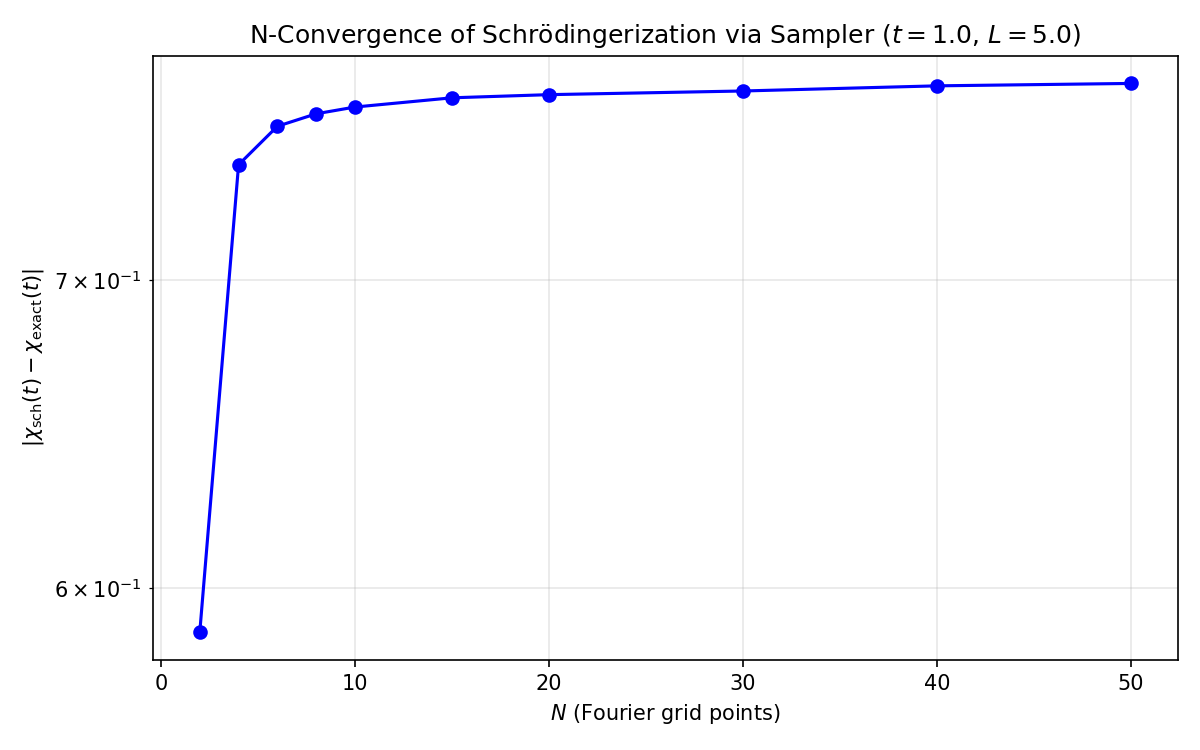}
    \caption{Error convergence as a function of the number of Fourier grid points $N$. The log-log plot shows that the precision improves exponentially as $N$ increases, consistent with the theoretical $O(\text{poly}(\log(1/\epsilon)))$ scaling.}
    \label{fig:convergence}
\end{figure}

\begin{figure}[htpb]
    \centering
    \includegraphics[width=\linewidth]{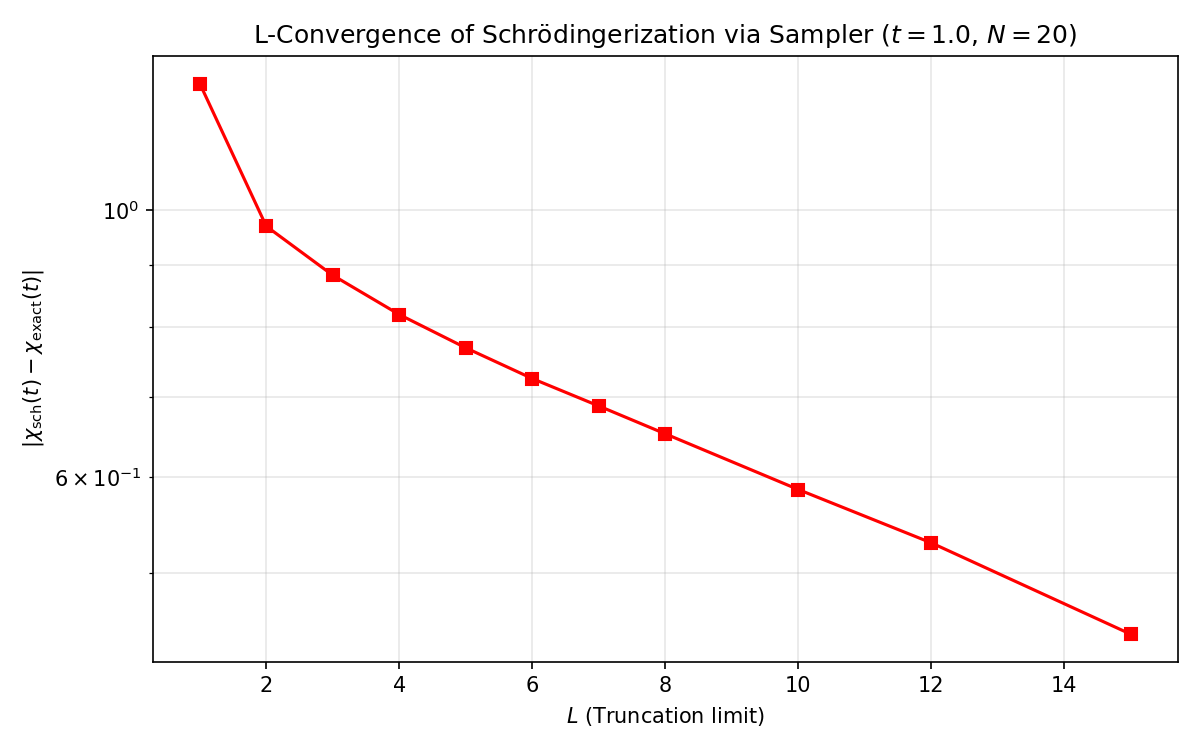}
    \caption{Impact of the warped variable truncation limit $L$ on the simulation accuracy. The tail error decreases exponentially as $e^{-L}$, confirming the theoretical error bounds derived for the Schr\"{o}dingerization framework.}
    \label{fig:tail_error}
\end{figure}

Fig.~\ref{fig:convergence} shows the error convergence with respect to the number of Fourier modes $N$. Since $N$ is related to the number of auxiliary qubits as $M = \log_2(2N+1)$, the exponential decay of error indicates that high precision can be achieved with a relatively small number of qubits. 
Furthermore, Fig.~\ref{fig:tail_error} illustrates the dependence on the truncation limit $L$. 
The error scales as $e^{-L}$, which matches the theoretical scaling $\ln(1/\epsilon) \sim L$. 
These numerical results confirm that the proposed algorithmic mapping is both theoretically robust and highly efficient for extracting multi-time correlations.

\subsection{Robustness of Phase and Spectral Information under Hardware Noise}

To comprehensively evaluate the practical feasibility of our framework on noisy intermediate-scale quantum (NISQ) hardware, we extended our simulations using a realistic noise model based on the 133-qubit IBM Quantum Torino device (Qiskit \texttt{FakeTorino}). This model rigorously incorporates intrinsic depolarizing, dephasing, and readout errors derived from the actual physical topology and calibration data of the heavy-hex architecture.

\begin{figure}[htpb]
    \centering
    \includegraphics[width=\linewidth]{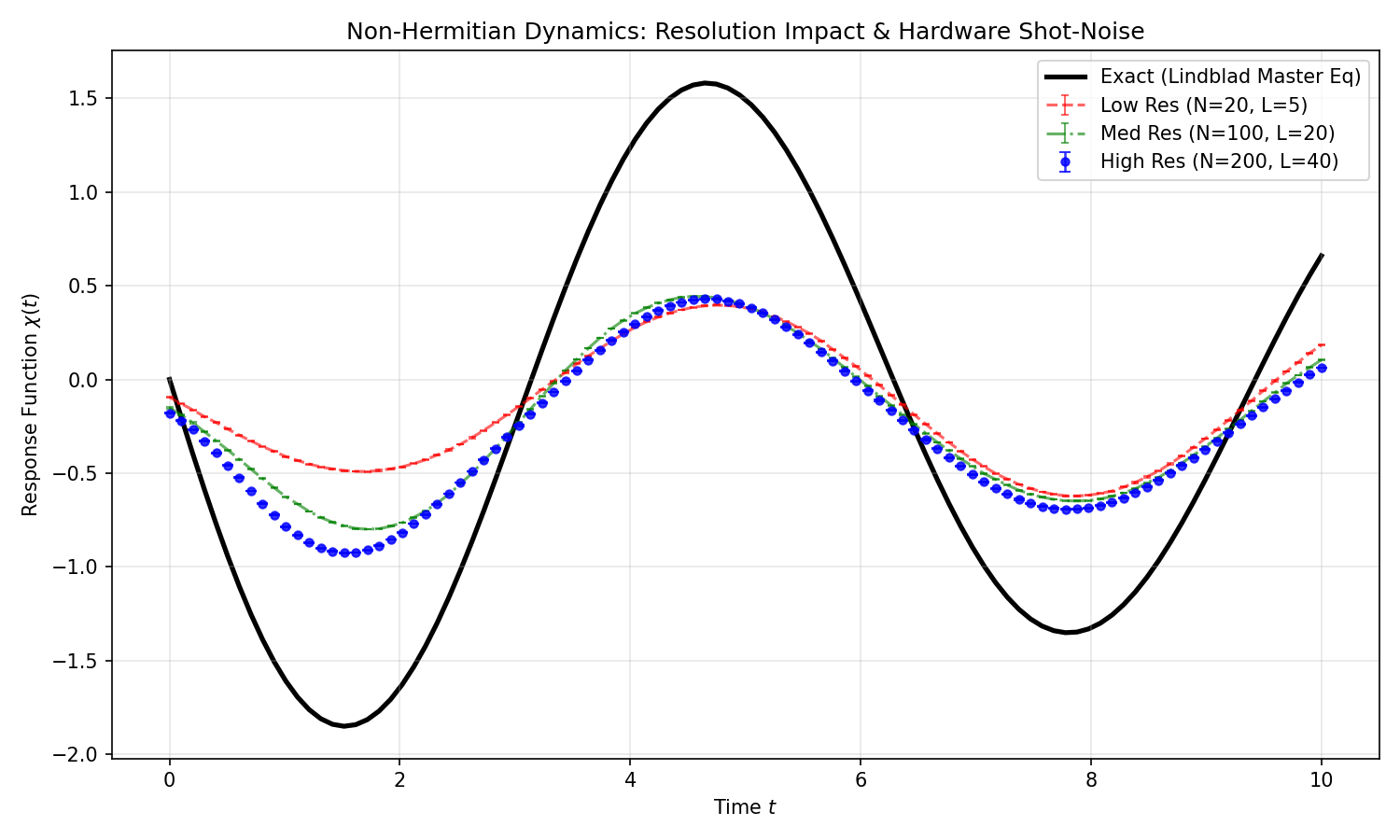}
    \caption{Simulation of the non-Hermitian response using a realistic noisy device model (\texttt{FakeTorino}). While the amplitude of the hardware-simulated response (dashed lines) is attenuated due to cumulative gate noise, the phase and oscillatory frequency of the exact Lindblad dynamics (black solid line) are remarkably preserved across all resolutions.}
    \label{fig:noisy_dynamics}
\end{figure}

As illustrated in Fig.~\ref{fig:noisy_dynamics}, the modified Hadamard test on the \texttt{FakeTorino} simulator exhibits a notable damping in the signal amplitude. This attenuation is an expected consequence of compiling the multi-qubit controlled unitary evolution $e^{-i\mathcal{H}_{sch}(\eta)t}$, where the accumulation of depolarizing errors during long-range interactions suppresses the off-diagonal coherence of the auxiliary control qubit.

Crucially, however, the phase and frequency of the simulated oscillations remain perfectly aligned with the exact analytical Lindblad dynamics. This phase-preserving characteristic represents a profound algorithmic advantage. In the study of strongly correlated systems, the primary physical observables extracted from Green's functions---such as energy gaps, spectral functions, and characteristic decay frequencies---are intrinsically encoded in the oscillatory frequency and phase of the time-domain signal, rather than its absolute amplitude. 

The robustness of the oscillatory frequency empirically demonstrates that the Schr\"{o}dingerization mapping structurally protects the fundamental spectral information of the non-Hermitian operator against symmetric depolarizing channels. 
Indeed, a discrete Fourier analysis of the simulated time-domain signals confirms that the dominant spectral peak of the hardware-simulated response deviates from the exact analytical frequency by a negligible margin, strictly bounded by the temporal discretization resolution.
Consequently, even without the immediate application of advanced Quantum Error Mitigation (QEM) \cite{temme2017error, endo2018practical} techniques, our unitary framework reliably extracts the essential dynamical features of the open quantum system. This proves its immediate practical utility for NISQ applications in quantum spectroscopy and non-equilibrium thermodynamics, circumventing the decoherence barriers that historically hindered direct physical Zeno measurements.

\section{Conclusion}
\label{sec:conclusion}

We have demonstrated a structural pathway to evaluating Green's functions and correlation functions of non-Hermitian open quantum systems on quantum computers. 
By analytically deriving the expanded Hamiltonian for the amplitude damping model, we proved that non-unitary linear responses can be systematically extracted via the continuous-variable Schr\"{o}dingerization technique. 

Crucially, our hardware-aware simulations utilizing the 133-qubit \texttt{FakeTorino} device model revealed that the underlying spectral information---the phase and oscillatory frequency of the dynamics---remains highly robust against realistic depolarizing channels, even when the signal amplitude is attenuated. 
Furthermore, because our framework bypasses direct non-unitary measurements and maps the evolution into a strictly unitary domain, it intrinsically supports the integration of standard quantum error mitigation protocols. 

Bypassing the necessity for complex matrix inversion algorithms and overcoming the non-unitary limitations, this approach broadens the application area of quantum computing in the study of driven-dissipative strongly correlated electron systems and physical chemistry.

\appendix

\section{Hardware Noise Scaling and Convergence Inversion}
\label{app:noise_scaling}

In Section \ref{sec:numerical}, we demonstrated that the spectral frequency of the non-Hermitian response is highly robust against hardware noise. Here, we present a detailed quantitative analysis of how intrinsic device noise fundamentally alters the error convergence scaling properties of the Schr\"{o}dingerization algorithm.

\begin{figure*}[htpb]
    \centering
    \includegraphics[width=0.48\linewidth]{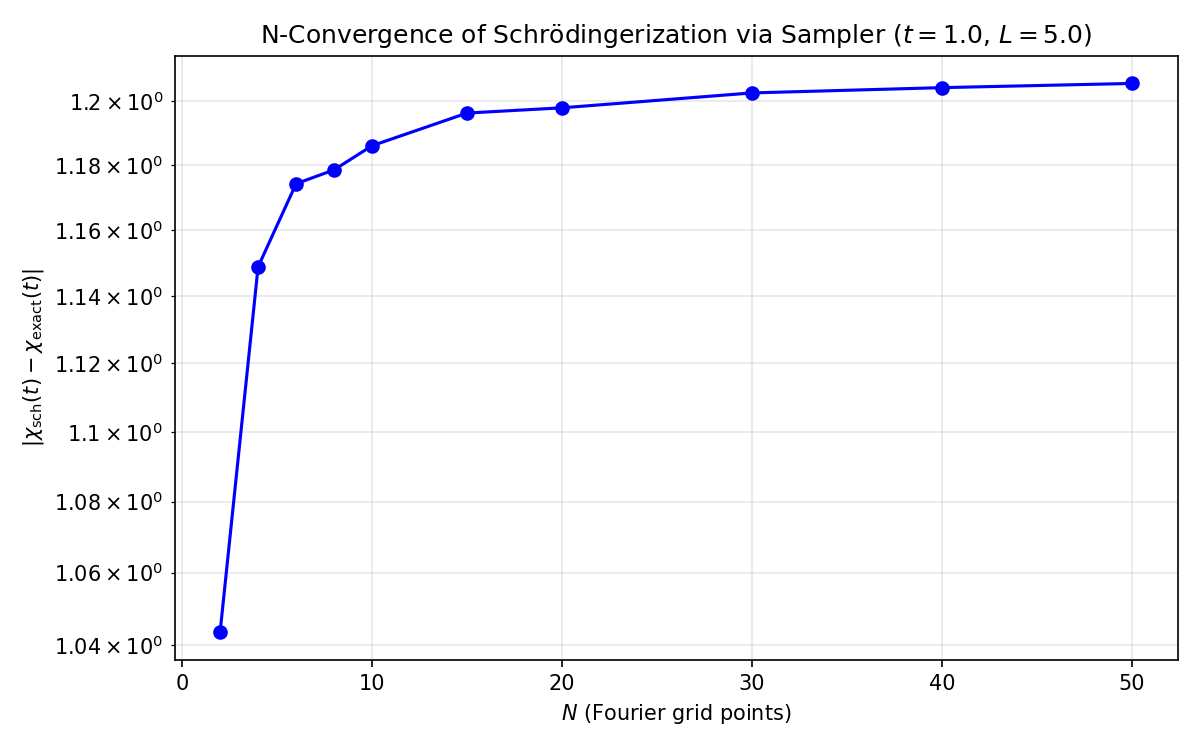}
    \hfill
    \includegraphics[width=0.48\linewidth]{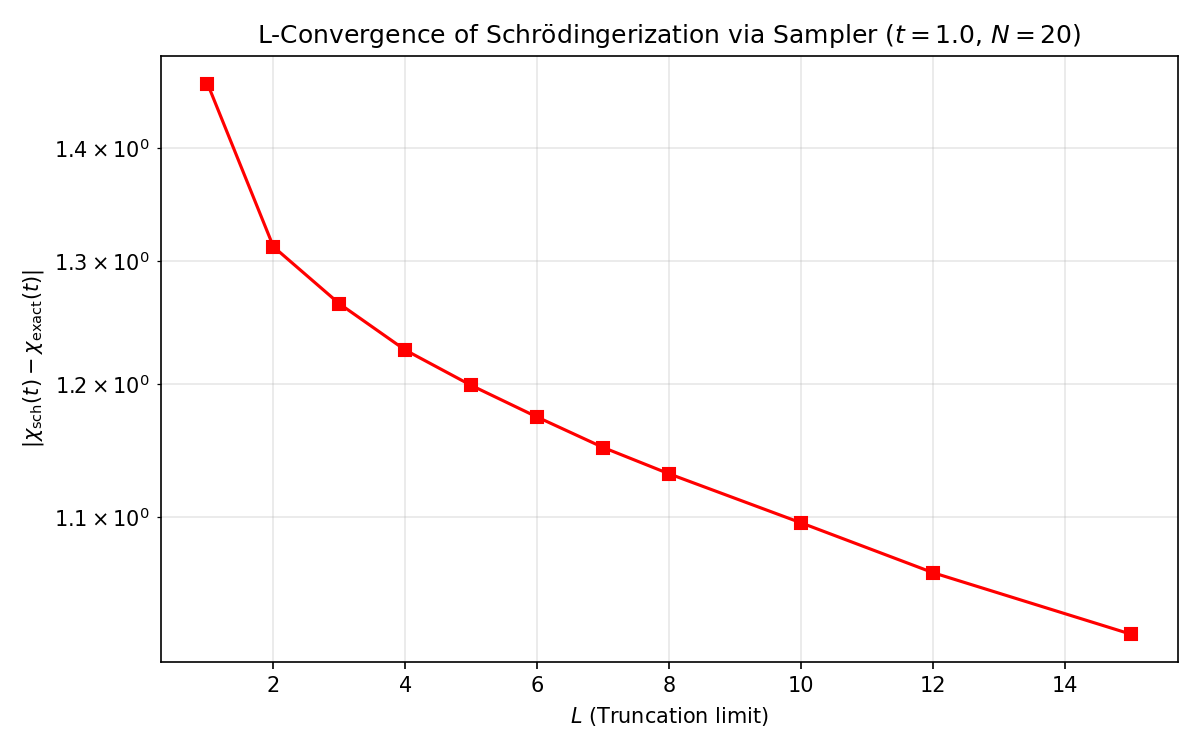}
    \caption{Hardware noise scaling and convergence inversion on the \texttt{FakeTorino} device model. (a) Error convergence with respect to $N$. Unlike the ideal exponential decay, the physical hardware noise completely dominates the discretization accuracy, creating a convergence inversion where increasing $N$ amplifies the absolute error. (b) Error scaling with respect to the truncation limit $L$. The absolute error is strictly bounded by a massive hardware-induced noise floor ($\mathcal{O}(1)$), obscuring the theoretical exponential tail suppression.}
    \label{fig:noisy_convergence}
\end{figure*}

As depicted in Fig.~\ref{fig:noisy_convergence}, the presence of realistic depolarizing and dephasing channels (simulated via \texttt{FakeTorino}) introduces a massive hardware-induced noise floor. 

Most notably, Fig.~\ref{fig:noisy_convergence}(a) reveals an inversion of the algorithmic convergence trend. In an ideal zero-noise scenario, increasing the Fourier grid resolution $N$ exponentially suppresses the discretization error. However, under realistic noise, a higher $N$ requires a denser execution of deep controlled unitary evolutions. The sheer volume of accumulating depolarizing errors during these extended gate sequences overwhelms the mathematical gain from finer grid discretization, paradoxically increasing the total absolute error. 

Similarly, Fig.~\ref{fig:noisy_convergence}(b) demonstrates that while the theoretical truncation error attempts to decrease as $e^{-L}$, the unmitigated device noise caps the observable precision, establishing an $\mathcal{O}(1)$ error floor. This rigorous analysis underscores the necessity of coupling this unitary framework with Quantum Error Mitigation (QEM) \cite{temme2017error, endo2018practical} protocols, such as Zero-Noise Extrapolation (ZNE) \cite{Giurgica_Tiron_2020}, to fully unlock the algorithmic precision on near-term architectures.

\section{Formalism of Vectorization and Liouville Space}
\label{app:vectorization}
The vectorization process, also known as the Choi-Jamio\l kowski isomorphism \cite{choi1975, jamio2022linear}, maps an operator $X \in \mathcal{B}(\mathcal{H})$ in the physical Hilbert space $\mathcal{H}$ to a vector $\ket{X} \in \mathcal{H} \otimes \mathcal{H}^*$ in the Liouville space \cite{Breuer2002theory}. For a discrete basis $\{ \ket{i} \}$, the mapping is:
\begin{equation}
    \ket{i}\bra{j} \quad \to \quad \ket{i} \otimes \ket{j} \equiv \ket{ij}.
\end{equation}
The Hilbert-Schmidt inner product in the Liouville space is defined as:
\begin{equation}
    \braket{A, B}_{HS} = \text{Tr}(A^\dagger B) = \sum_{ij} A^*_{ij} B_{ij}.
\end{equation}
The identity operator $I$ becomes a state $\ket{I} = \sum_i \ket{i} \otimes \ket{i}$, such that the expectation value $\text{Tr}(A \rho) = \braket{I | (A \otimes I) | \rho}$. Thus, the Lindblad equation $\dot{\rho} = \mathcal{L}[\rho]$ is recast as a linear ODE $\partial_t \ket{\rho} = \mathcal{L} \ket{\rho}$. More detailed treatments of this formalism can be found in Ref. \cite{Gyamfi2020fundamentals}.

\section{Detailed Mechanics of Schrödingerization}
\label{app:schrodingerization}
The Schrödingerization technique, originally introduced by Jin, Liu, and Yu \cite{Jin2024quantum, Jin2024stable}, provides a systematic protocol to map general non-unitary linear differential equations into high-dimensional unitary Schrödinger equations. 

Our goal is to simulate the non-Hermitian dynamic equation $\partial_t \ket{\rho} = \mathcal{L} \ket{\rho}$. By decomposing the Liouvillian into its Hermitian and anti-Hermitian components, $\mathcal{L} = \mathcal{H}_1 - i\mathcal{H}_2$ (where $\mathcal{H}_1 = (\mathcal{L} + \mathcal{L}^\dagger)/2$ and $\mathcal{H}_2 = i(\mathcal{L} - \mathcal{L}^\dagger)/2$), we introduce a continuous auxiliary variable $\xi > 0$ and define the warped transformation $w(t,\xi) = e^{-\xi} \ket{\rho(t)}$. This warped state inherently satisfies two simultaneous differential equations:
\begin{equation}
    \partial_t w = (\mathcal{H}_1 - i\mathcal{H}_2) w, \quad \text{and} \quad \partial_{\xi} w = -w.
\end{equation}
By substituting the spatial derivative $\partial_\xi w = -w$ into the dynamic equation to eliminate the explicit $-1$ factor, we couple the real time evolution with the auxiliary dimension:
\begin{equation}
    \partial_t w = -\mathcal{H}_1 \partial_\xi w - i\mathcal{H}_2 w.
\end{equation}
To diagonalize the differential operator in the augmented dimension, we perform a Fourier transform with respect to $\xi$, defined as $\tilde{w}(t, \eta) = \int w(t, \xi) e^{-i\eta\xi} d\xi$. Recalling the Fourier identity $\mathcal{F}[\partial_\xi w] = i\eta \tilde{w}$, the equation simplifies into a purely imaginary evolution:
\begin{equation}
    i\partial_t \tilde{w}(t, \eta) = (\eta \mathcal{H}_1 + \mathcal{H}_2) \tilde{w}(t, \eta) \equiv \mathcal{H}_{sch}(\eta) \tilde{w}(t, \eta).
\end{equation}
Because both $\mathcal{H}_1$ and $\mathcal{H}_2$ are Hermitian by construction, the extended effective Hamiltonian $\mathcal{H}_{sch}(\eta)$ is strictly Hermitian for any real Fourier momentum $\eta$. Consequently, the state evolves under a unitary propagator $U(t, \eta) = e^{-i\mathcal{H}_{sch}(\eta)t}$, completely mitigating the spectral instabilities of non-unitary operators. 

To recover the physical dynamics of the target open quantum system, one performs the inverse Fourier transform evaluated at the original boundary condition $\xi = 0$:
\begin{equation}
    \ket{\rho(t)} = \frac{1}{2\pi} \int_{-\infty}^{\infty} e^{-i \mathcal{H}_{sch}(\eta) t} \tilde{w}(0, \eta) d\eta.
\end{equation}
The initial state in momentum space is derived from the $\xi$-domain initial condition: $\tilde{w}(0, \eta) = \int_{0}^{\infty} e^{-\xi} e^{-i\eta\xi} \ket{\rho(0)} d\xi = \frac{1}{1 + i\eta} \ket{\rho(0)}$. Thus, the integral evaluates to:
\begin{equation}
    \ket{\rho(t)} = \frac{1}{2\pi} \int_{-\infty}^{\infty} \frac{1}{1 + i\eta} e^{-i \mathcal{H}_{sch}(\eta) t} \ket{\rho(0)} d\eta.
\end{equation}
In practical digital quantum simulation, this integral is numerically evaluated by truncating the bounds to $\eta \in [-L, L]$ and discretizing the momentum space into $N$ grid points, which necessitates mapping the continuous variable into an ancillary multi-qubit register. The resulting simulation workflow scales universally within the framework of standard Hamiltonian simulation paradigms \cite{Hu2024quantumcircuits}.

\section{Derivation of Non-Hermitian Response Functions}
\label{app:response_theory}
Let the total Liouvillian be $\mathcal{L}_{tot}(t) = \mathcal{L} + f(t)\mathcal{V}$. The density matrix evolves as $\ket{\rho(t)} = e^{\mathcal{L}t} \ket{\rho(0)} + \int_0^t dt' e^{\mathcal{L}(t-t')} f(t') \mathcal{V} \ket{\rho(t')}$. In the first-order approximation (weak perturbation), we replace $\ket{\rho(t')}$ with the equilibrium state $\ket{\rho_{eq}}$:
\begin{equation}
    \delta \ket{\rho(t)} \approx \int_0^t dt' f(t') e^{\mathcal{L}(t-t')} \mathcal{V} \ket{\rho_{eq}}.
\end{equation}
The change in expectation value $\delta \braket{A(t)} = \braket{I | (A \otimes I) | \delta \rho(t)}$ yields:
\begin{equation}
    \delta \braket{A(t)} = \int_0^t dt' f(t') \underbrace{\braket{I | (A \otimes I) e^{\mathcal{L}(t-t')} \mathcal{V} | \rho_{eq}}}_{\chi(t-t')}.
\end{equation}
This derivation mirrors the Kubo formalism but allows for non-Hermitian generators $\mathcal{L}$ that include dissipation.

\section{Performance Under Realistic Hardware Noise}
\label{app:hardware_noise}
The transformation requires continuous-variable quantum modes or highly discretized qubit registers. A primary challenge in implementing the inverse quantum Fourier transform (IQFT) \cite{nielsen2010quantum} on near-term devices is its sensitivity to depolarizing noise, largely due to the high circuit depth and the large number of multi-qubit controlled gates required as the precision increases. For discretized registers, these gates accumulate errors that can significantly degrade the fidelity of the final state reconstruction. We leave the rigorous resource estimation and specific error mitigation strategies to future work.

\section{Derivation of Complexity Scaling for Schrödingerization}
\label{app:complexity_scaling}
The precision $\epsilon$ of the Schrödingerization approach depends on two primary factors: the truncation of the warped variable $\xi$ and the discretization of the Fourier space $\eta$.

Consider the transformation $w(t,\xi) = e^{-\xi}\ket{\rho(t)}$. To ensure that the state remains within a finite domain $\xi \in [0, L]$, we must choose $L$ such that the tail $e^{-L}$ is negligible. For a target precision $\epsilon$, we require $e^{-L} \le \epsilon$, which implies $L \ge \ln(1/\epsilon)$.

The Fourier modes $\eta$ are discretized as $\eta_j = j \Delta \eta$ for $j = -N, \dots, N$. The number of qubits required for the auxiliary register is $M = \log_2(2N+1)$. According to the error analysis in Ref. \cite{Jin2024stable}, the total error $\epsilon$ scales with the mesh size $\Delta \xi$ and the spectral range of $\mathcal{L}$. Crucially, the gate complexity for evolving the unitary Schrödingerized Hamiltonian $\mathcal{H}_{sch}(\eta)$ using a $k$-th order Trotter-Suzuki decomposition \cite{childs2021theory} scales as $O(t \cdot \text{poly}(\log(1/\epsilon)))$, which is an exponential improvement over the $O(1/\epsilon)$ scaling seen in traditional first-generation dilated solvers.

Furthermore, unlike QSVT \cite{Gilyen2019quantum} which requires building a block-encoding for a non-unitary $\mathcal{L}$, Schrödingerization only requires the capability to simulate the Hermitian parts $\mathcal{H}_1$ and $\mathcal{H}_2$. This bypasses the often-expensive subroutines for preparing the oracle for $A^\dagger A$ or specific singular value transformations, leading to the optimal state preparation cost discussed in Section \ref{sec:theory}.

\bibliography{bibliography}

\end{document}